%% file: main.tex
\documentclass[lettersize,journal]{IEEEtran}
\usepackage{amsmath,amsfonts}
\usepackage{algorithmic}
\usepackage{algorithm}
\usepackage{array}
\usepackage[caption=false,font=normalsize,labelfont=sf,textfont=sf]{subfig}
\usepackage{textcomp}
\usepackage{stfloats}
\usepackage{siunitx}
\usepackage{url}
\usepackage{physics}
\usepackage{verbatim}
\usepackage{graphicx}
\usepackage{stmaryrd}
\usepackage{cite}
\usepackage{soul,color}
\usepackage[colorlinks=true,breaklinks=true,allcolors=blue]{hyperref}
\usepackage[capitalise,nameinlink]{cleveref}

\begin{document}

\title{Towards scalable cryogenic quantum dot biasing using memristor-based DC sources}

\author{
    \IEEEauthorblockN{Pierre-Antoine~Mouny$^{1,2,3}$, Raphaël~Dawant$^{1,2,3}$, Patrick~Dufour$^{1,2,3}$, Matthieu~Valdenaire$^{1,2,3}$, Serge~Ecoffey$^{1,2}$, Michel~Pioro-Ladrière$^{2,3}$, Yann~Beilliard$^{1,2,3}$ and Dominique~Drouin$^{1,2,3}$}
     \\\footnotesize$^1$\textit{Institut Interdisciplinaire d’Innovation Technologique (3IT), Université de Sherbrooke, Sherbrooke, Québec J1K 0A5, Canada}
     \\\footnotesize$^2$\textit{Laboratoire Nanotechnologies Nanosystèmes (LN2) – CNRS UMI-3463 – 3IT, Sherbrooke, Québec J1K 0A5, Canada}
     \\\footnotesize$^3$\textit{Institut quantique (IQ), Université de Sherbrooke, Sherbrooke, Québec J1K 2R1, Canada}
     \\\footnotesize\textbf{CORRESPONDING AUTHOR: P.-A. MOUNY (e-mail: pierre-antoine.mouny@usherbrooke.ca)}}

\markboth{}%
{Mouny \MakeLowercase{\textit{et al.}}:Towards scalable cryogenic quantum dot biasing using memristor-based DC sources}


\maketitle

\begin{abstract}
Cryogenic memristor-based DC sources offer a promising avenue for in situ biasing of quantum dot arrays. In this study, we present experimental results and discuss the scaling potential for such DC sources. We first demonstrate the operation of a commercial discrete operational amplifier down to \SI{1.2}{\kelvin} which is used on the DC source prototype. Then, the tunability of the memristor-based DC source is validated by performing several \SI{250}{\milli\volt}-DC sweeps with a resolution of \SI{10}{\milli\volt} at room temperature and at \SI{1.2}{\kelvin}. Additionally,  the DC source prototype exhibits a limited output drift of $\approx\SI{1}{\micro\volt\per\second}$ at \SI{1.2}{\kelvin}. This showcases the potential of memristor-based DC sources for quantum dot biasing. Limitations in power consumption and voltage resolution using discrete components highlight the need for a fully integrated and scalable complementary metal–oxide–semiconductor-based (CMOS-based) approach. To address this, we propose to monolithically co-integrate emerging non-volatile memories (eNVMs) and \SI{65}{\nano\metre} CMOS circuitry. Simulations reveal a reduction in power consumption, down to \SI{10}{\micro\watt} per DC source and in footprint.  This allows for the integration of up to one million eNVM-based DC sources at the \SI{4.2}{\kelvin} stage of a dilution fridge, paving the way for near term large-scale quantum computing applications.
\end{abstract}

\begin{IEEEkeywords}
Memristors, Cryogenic electronics, Quantum dots (QDs)
\end{IEEEkeywords}

\input{01_Introduction}

\input{02_Prototype}
\input{03_Cryo_Amp}

\input{04_Exp_Setup}
\input{05_Measurement}
\input{06_Scaling}

\input{07_Conclusion}

\input{08_Others}
\bibliographystyle{unsrt}
\bibliography{references}
\clearpage

\end{document}

%% file: 01_Introduction.tex
\section{Introduction}

\IEEEPARstart{Q}{uantum} computing promises breakthrough applications in a variety of fields, including chemistry \cite{AspuruGuzik2005}, finance \cite{Ors2019} and climate \cite{berger2021quantum}. Several physical platforms have been proposed to realize quantum computers, ranging from superconducting circuits \cite{Nakamura1999, Arute2019} to trapped ions \cite{Cirac1995, Leibfried2003}. Among these candidates, silicon quantum dots (QDs) benefit from a nanometric qubit pitch \cite{Bedecarrats2021}, a long coherence time \cite{Veldhorst2014} compared to traditional superconducting qubits, and a temperature of operation up to \SI{4.2}{\kelvin} \cite{Petit2020, Yang2020, Camenzind2022}. Additionally, QDs appear to be a highly scalable technology thanks to their compatibility with industrial semiconductor fabrication processes \cite{Stuyck2021, Zwerver_2022}, demonstrating their potential for industrial-scale production and monolithic co-integration with control electronics \cite{XuSwitchedCapa2020}. Yet, millions of physical qubits compatible with quantum error correction protocols will be required to unlock these promised applications \cite{Fowler_2012}. Recently, notable advancements have been made, such as enhancements in gate fidelity \cite{Xue2022}, early scalable QDs matrices \cite{Lawrie2020, Borsoi2023}, and large-scale fabrication of qubits \cite{Ansaloni2020, Geyer2021}. While these developments contribute to scaling to millions of physical qubits, the linear approach used to control these QDs emerges as a bottleneck, impeding the seamless integration of an increasing number of qubits. To address this challenge, concurrent scaling of control electronics becomes imperative to ensure the scalability of the QD control method. 
Early scaling propositions have explored the avenue of cryo-complementary metal–oxide–semiconductor (cryo-CMOS) technology \cite{Charbon2016, Patra2018}. In particular, mixed-signal solutions have emerged as a principal focus to perform state manipulation and readouts with cryogenic digital-to-analog converters (DACs) to generate pulses \cite{Xue2021, Pauka2021}, cryogenic flip-flop memory to store measurement protocols \cite{Geck2019}, and amplifier/mixer circuits \cite{Prabowo2021}. Cryogenic DACs with monolithically integrated switched-capacitors have also been proposed to perform in situ biasing of QD gates \cite{XuSwitchedCapa2020, ruffino2021integrated}. The required biasing voltages of the QDs are stored in the charge of the capacitors conceptually close to dynamic random access memories (DRAMs) \cite{Geck2019}. This approach benefits from ultra-low power dissipation in the few tens of picowatts but requires periodical refreshment of the capacitor charge due to leakage current, approximately every \SI{10}{\micro\second}-\SI{1}{\milli\second} \cite{Geck2019, ruffino2021integrated}, to maintain the integrity of the biasing voltage. To avoid this volatility, which will be problematic when scaling up QD-based systems, a memristor-based biasing circuit has been proposed. This circuit exploits the non-volatility of TiO$_\textrm{x}$ memristors at the cost of power dissipation on the order of a few milliwatts \cite{Mouny2023TED}, which could be reduced to tens of microwatts by using integrated circuits. The behavior of TiO$_\textrm{x}$-based memristors has been widely studied at cryogenic temperatures, demonstrating DC resistive switchings \cite{Pickett2011, Beilliard2020} and analog programming with up to 4-bit memristors \cite{Mouny2023TED, MounyAPL2023}. However, the concept of memristor-based DC sources is yet to be experimentally demonstrated at cryogenic temperatures.\\

In this paper, we investigate the cryogenic DC behavior of a transimpedance amplifier (TIA) based on a commercial operational amplifier (OpAmp) AD8605 between \SI{1.2}{\kelvin} and \SI{300}{\kelvin}. We then propose a prototype of a memristor-based DC source using the cryo-compatible AD8605 OpAmp. This circuit is characterized at \SI{1.2}{\kelvin} and at room temperature which serves as a performance benchmark. We perform DC sweeps with a voltage range of \SI{250}{\milli\volt} and a \SI{10}{\milli\volt} resolution. Additionally, we study the stability of the output voltage ensuring that it does not change over time due to memristor resistance drift. 
Finally, we discuss the scalability of this prototype and propose an alternative design to reduce the overall power consumption and footprint. This architecture supposes near-perfect memristors co-integrated with \SI{65}{\nano\metre} CMOS technology to bias the gates of a silicon quantum dot cooled to \SI{4.2}{\kelvin}.

%% file: 02_Prototype.tex
\section{Memristor-based DC source prototype}
\begin{figure}
\begin{minipage}[h]{1\columnwidth}
  \includegraphics[width=0.9\linewidth]{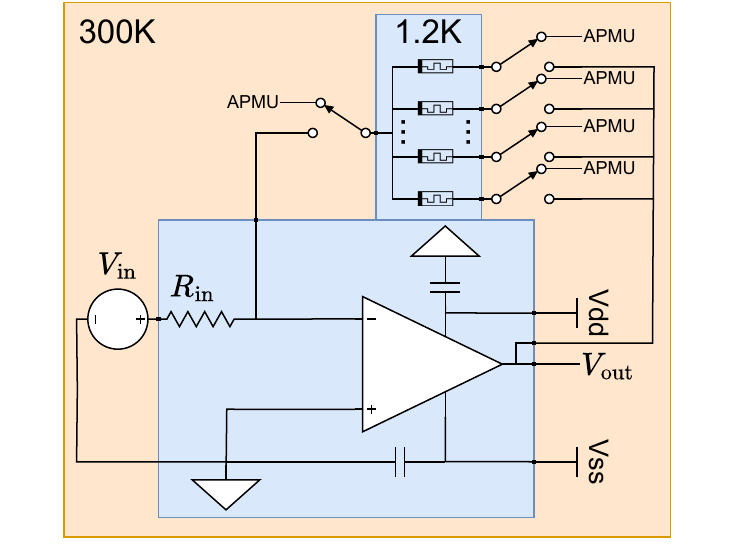}
  \caption{\label{fig:concept}\textbf{Schematic view of the memristor-based DC source prototype.} The required interconnects between cryogenic temperatures and room temperature electronics are shown as small black squares. Analog switches are used to connect the memristors to APMUs or to short them to provide resistive feedback to the OpAmp.}
\end{minipage}\hfill
\end{figure}

Earlier work conceptualized a memristor-based DC source compatible with cryogenic temperatures \cite{Mouny2023TED}. This concept uses a single OpAmp placed in a TIA circuit configuration. The resistive feedback of the TIA is achieved using either a single memristor or multiple memristors in parallel placed in the feedback loop of the OpAmp (see \cref{fig:concept}). The output voltage of the memristor-based DC source can be tuned by changing the feedback resistance. This is achieved by individually programming each memristor placed in the feedback loop of the OpAmp, effectively building a programmable gain amplifier (PGA) whose output voltage, $V_\textrm{out}$, depends on the variable gain $G_v$:
\begin{equation}
    \abs{V_\textrm{out}} = \frac{R_\textrm{mem}}{R_\textrm{in}}\times V_\textrm{in}
\end{equation}
where $R_\textrm{mem}$ is the total feedback resistance introduced by the memristors and is given by:
\begin{equation}
    R_\textrm{mem} = (\sum_{i}^{N}G_{i})^{-1}
\end{equation}
where $i$ represents the index of the i-th memristor, $G_i$ is the conductance of the i-th memristor in the feedback loop, and $N$ the number of memristors in the feedback loop e.g., $N=4$ in \cref{fig:concept}.\\

The memristors in the feedback loop can be individually programmed by using analog switches placed at room temperature; $N+1$ switches are needed as a common top electrode is used for all memristors. The analog switch at the common top electrode allows it to connect to an arbitrary pulse measurement unit (APMU) which can send pulses to program the memristors or connect to the feedback loop of the OpAmp. Meanwhile, each memristor's bottom electrode is connected to an analog switch, enabling the grounding of the bottom electrode with an APMU for memristor programming, or shorting all bottom electrodes to create a feedback resistance with the memristors in parallel (see \cref{fig:concept}).\\

The cryogenic compatibility of this prototype presents two primary challenges. Firstly, finding a commercial operational amplifier capable of functioning at deep cryogenic temperatures down to \SI{1.2}{\kelvin}. Secondly, validating the cryogenic compatibility of the memristors. Cryo-compatible memristors have already been demonstrated with different oxides such as HfO$_\textrm{x}$ \cite{Voronkovskii2019, Fang2015}, HfZnO \cite{Lan2023}, TaO$_\textrm{x}$ \cite{Zhang2014} and TiO$_\textrm{x}$ \cite{Pickett2011, Beilliard2020, MounyAPL2023}. In the meantime, a few commercial amplifiers have been characterized at cryogenic temperatures down to \SI{4.2}{\kelvin} demonstrating cryo-compatibility. These OpAmps include the TLC271 from Texas Instruments \cite{Proctor2015}, the TLV271 from onsemi, the AD8601 and AD8605 from Analog Devices \cite{Homulle2019}. Due to its low power consumption, it was decided to investigate the cryo-compatibility of AD8605 in the next section .

%% file: 03_Cryo_Amp.tex
\begin{figure*}[htbp]
\centering
  \includegraphics[width=0.95\linewidth]{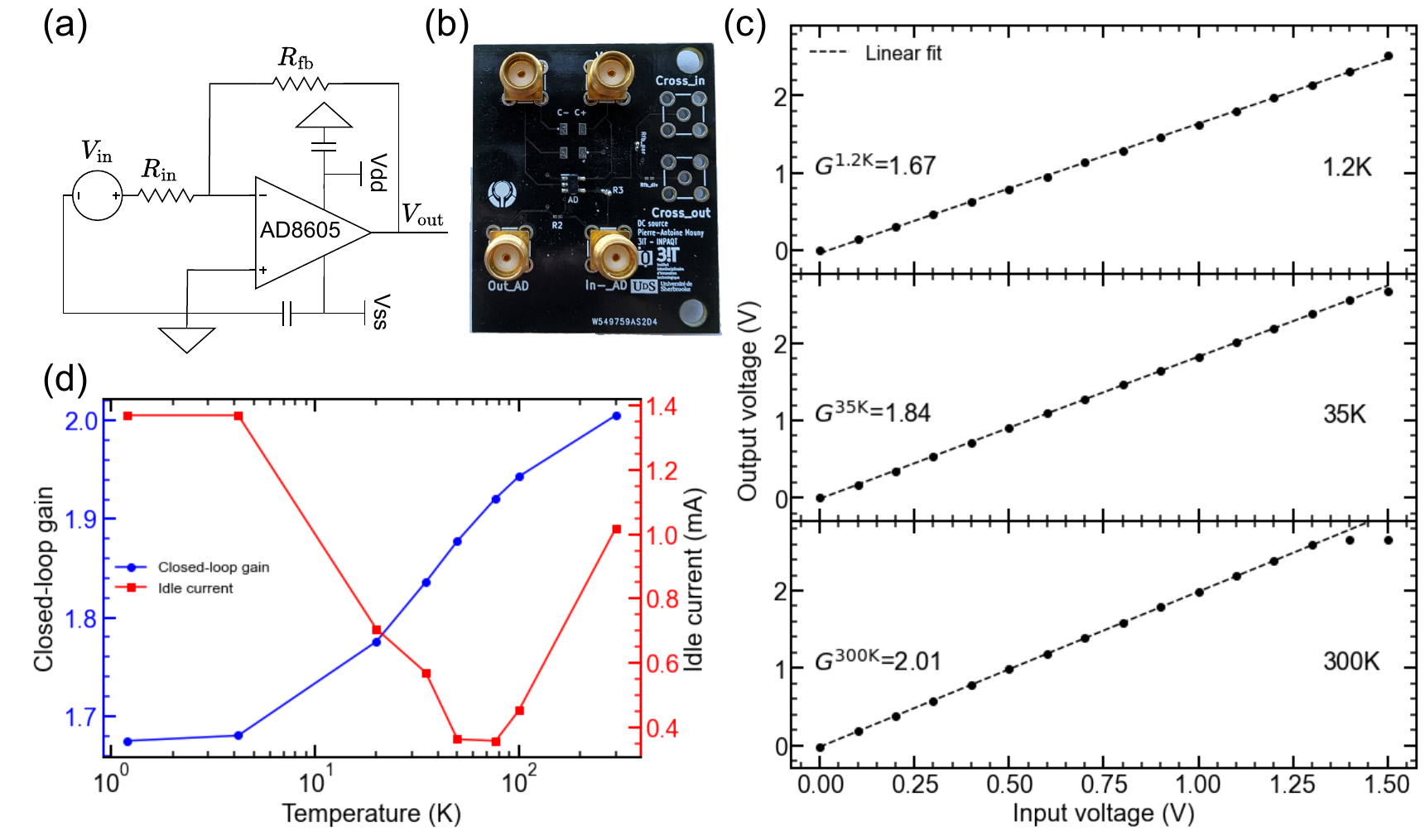}

\caption{\textbf{Cryogenic characterization of the AD8605 OpAmp} \textbf{a.} Schematic of the cryogenic TIA circuit based on the AD8605 OpAmp. $R_\textrm{fb}/R_\textrm{in}$ define the gain of the TIA with $R_\textrm{fb}=\SI{2}{\kilo\ohm}$ and $R_\textrm{in}=\SI{1}{\kilo\ohm}$. \textbf{b.} FR-4 PCB implementation of the AD8605 TI circuits. \textbf{c.} DC characterization of the TIA behavior at \SI{1.2}{\kelvin}, \SI{35}{\kelvin} and \SI{300}{\kelvin}. \textbf{d.} Fitted closed-loop gain (blue) and idle current (red) of the AD8605 OpAmp from \SI{1.2}{\kelvin} to \SI{300}{\kelvin}. The idle current is defined as the current consumption for $V_\textrm{in}=\SI{0}{\volt}$}
\label{fig:AD8605_carac}
\end{figure*}

\section{Characterization of the cryogenic amplifier}\label{cryo_amp}
 The AD8605 OpAmp is tested in a typical TIA configuration, as depicted in \cref{fig:AD8605_carac}a, to match the prototype presented in \cref{fig:concept}. This circuit is implemented on a 2-layer $46\times\SI{39}{\milli\metre}$ FR-4 PCB (see \cref{fig:AD8605_carac}b). The PCB is assembled with two resistors: $R_\textrm{in}=\SI{1}{\kilo\ohm}$ and $R_\textrm{fb}=\SI{2}{\kilo\ohm}$; along with two \SI{1}{\micro\farad} decoupling capacitors to limit power supply noise. This PCB is placed in the \SI{1}{\kelvin}-pot of an ICE Oxford DRY ICE cryostat allowing to cool down the TIA PCB to \SI{1.2}{\kelvin}. Local heating can be applied to the \SI{1}{\kelvin}-pot of the cryostat to increase the temperature of the PCB from \SI{1.2}{\kelvin} to \SI{300}{\kelvin}. A dual voltage supply of $\pm\SI{2.7}{\volt}$ is applied to the AD8605 OpAmp using two Stanford Research Systems DC205 high-precision DC sources. The supply voltages are maintained during the cooldown of the PCB. The supply current drawn by the AD8605 OpAmp is measured by placing a Keysight 34461A digital multimeter in series with the supply DC sources. This allows for the measurement of the power consumption of the cryogenic amplifier and the estimate of its power dissipation. The output voltage of the cryogenic amplifier is measured with a Keysight DSOX3014T oscilloscope. Four high-frequency coaxial copper lines are used to route the different signals in and out of the cryostat.\\

The DC behavior of the AD8605 TIA circuit is measured by sweeping the input voltage, $V_\textrm{in}$, from \SI{0}{\volt} to \SI{1.5}{\volt}. \cref{fig:AD8605_carac}c shows the results at \SI{1.2}{\kelvin}, \SI{35}{\kelvin} and room temperature, which serves as a benchmark DC behavior. The AD8605 shows a linear TIA-like behavior at both \SI{1.2}{\kelvin} and \SI{35}{\kelvin}, validating the cryo-compatibility of the AD8605 OpAmp. However, the gain of the AD8605 OpAmp is observed to decrease at lower temperatures. This trend is validated by fitting the gain of the TIA circuit at multiple cryogenic temperatures (see ~\cref{fig:AD8605_carac}c). The gain of the OpAmp exhibit a logarithmic increase with temperatures between \SI{4.2}{\kelvin} and \SI{300}{\kelvin}. Below \SI{4.2}{\kelvin}, the gain of the OpAmp plateaus roughly at 1.68. This gain decrease could partly be explained by $R_\textrm{in}$ being a thick film resistor whose resistance increases at lower temperatures \cite{Patterson_2001}.\\

Additionally, the current consumption of the AD8605 OpAmp is measured at $V_\textrm{in}=\SI{0}{\volt}$ with a \SI{50}{\ohm} load. This current is usually named idle current. As \cref{fig:AD8605_carac}c depicts, the idle current of the AD8605 OpAmp is larger below \SI{4.2}{\kelvin} and plateaus at \SI{1.4}{\milli\ampere}. Then, the idle current decreases with increasing temperature up to \SI{77}{\kelvin} where it reaches a lower bound of \SI{350}{\micro\ampere}. Finally, it increases with temperature reaching \SI{1}{\milli\ampere} at room temperature. The AD8605 OpAmp consumes 40\% more current at \SI{1.2}{\kelvin} than at room temperature for the same dual supply voltage of $\pm\SI{2.7}{\volt}$, leading to a power consumption of roughly \SI{4}{\milli\watt}. While this power consumption is too large for scaling up the memristor-based DC source concept proposed by Mouny et al. \cite{Mouny2023TED}, its cryo-compatibility makes it the perfect candidate for prototyping the memristor-based DC source.

%% file: 04_Exp_Setup.tex
\begin{figure*}[htbp]
\centering
  \includegraphics[width=0.95\linewidth]{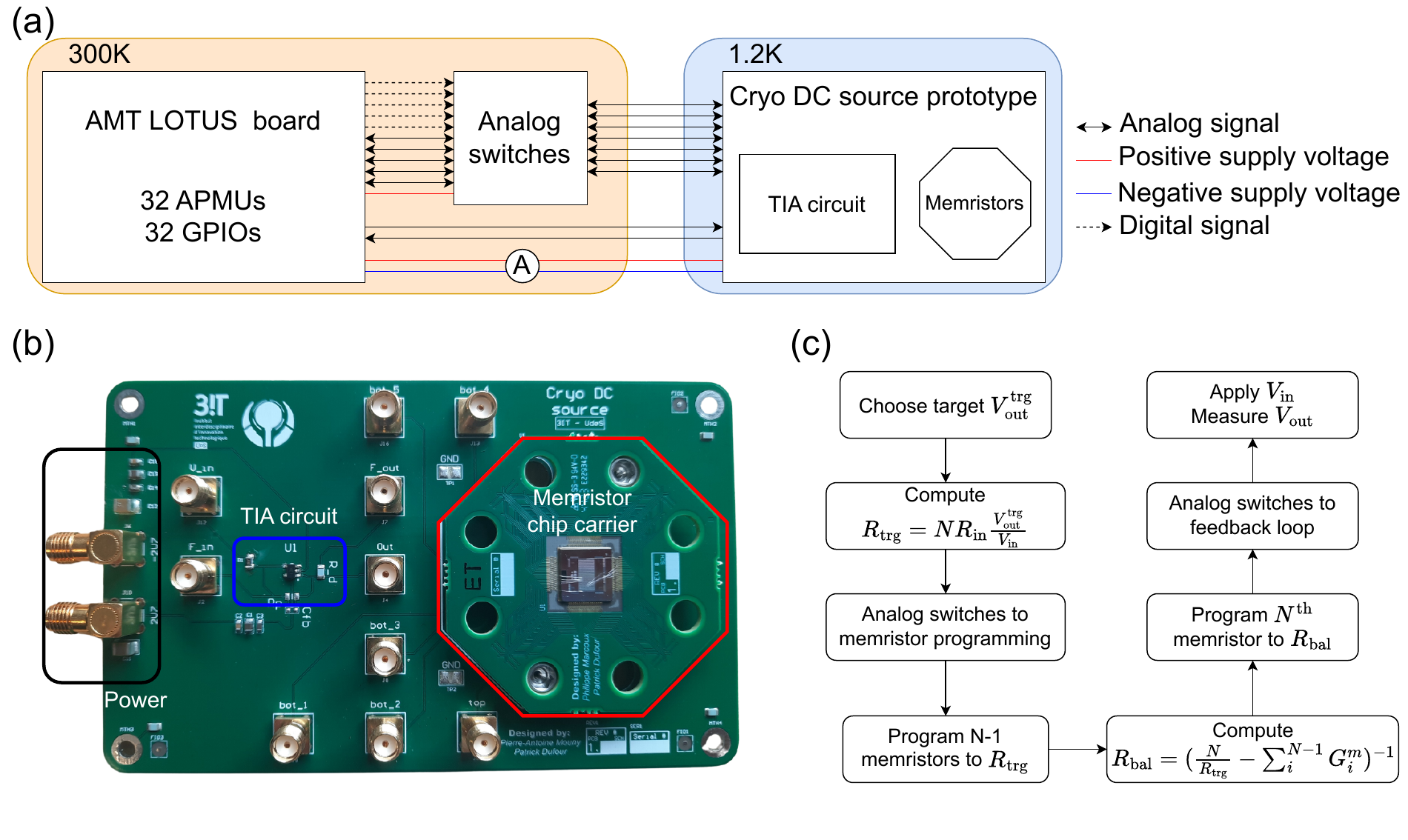}

\caption{\textbf{Experimental setup for the cryogenic memristor-based DC source.} \textbf{a.} Block schematic of the experimental setup used to validate the cryogenic memristor-based DC source. \textbf{b.}  4-layer FR-4 PCB implementation of the memristor-based DC source prototype. A custom chip carrier with a wire-bonded memristor chip is connected to the prototype PCB. \textbf{c.} Flowchart of the DC source prototype programming.}
\label{fig:Proto_exp_setup}
\end{figure*}
\section{Prototype experimental methods}
Having established the operation of the AD8605 OpAmp at cryogenic temperatures in the preceding section, we detail the setup used to experimentally demonstrate the memristor-based DC source prototype. \cref{fig:Proto_exp_setup}a shows a block schematic of the experimental setup which consists of three main blocks: a control platform, the analog switches, and the DC source prototype. The control platform is a LOTUS board from Advanced MicroTesting \cite{AMT}  that provides 32 asynchronous APMUs and 32 General Purpose Inputs/Outputs (GPIOs). A mezzanine board with Vishay Siliconix DG4053 analog switches is fabricated to either program the memristors or connect them to the feedback loop of the TIA circuit. Finally, a 4-layer FR-4 PCB is fabricated to co-integrate the memristor chip and the TIA circuit tested previously in Section \ref{cryo_amp} (see \cref{fig:Proto_exp_setup}b). The $R_\textrm{in}$ resistor is replaced with a \SI{3}{\kilo\ohm}-metallic film resistor to ensure compatibility with the memristor resistances, while $R_\textrm{fb}$ is replaced by a memristor feedback resistance as depicted by \cref{fig:concept}. \\

The memristor chip is glued to a custom chip carrier. A line of two memristors sharing a common top electrode is wedge-bonded with aluminum wires to the chip carrier using a TPT HB10 semi automatic wire bonder. This number of memristor yields to the smallest form factor memristor-based DC source required for the demonstration. The memristor chip is fabricated with an etch-back process described in Ref.\cite{Dawant2024}. Similar devices have been tested at cryogenic temperatures down to \SI{4.2}{\kelvin} and have demonstrated analog programmability enabled by cryogenic reforming \cite{MounyAPL2023}.
This chip carrier is connected to the cryogenic DC source PCB using mezzanine connectors.\\

The output voltage of the memristor-based DC source can be tuned by following the programming flowchart depicted in \cref{fig:Proto_exp_setup}c. Initially, a target output voltage, $V_\textrm{out}^\textrm{trg}$, is chosen which introduces a common target resistance $R_\textrm{trg}$ for the feedback memristors to be programmed. This resistance state is given by:
\begin{equation}
    R_\textrm{trg}=NR_\textrm{in}\frac{V_\textrm{out}^\textrm{trg}}{V_\textrm{in}}
\end{equation}
where $N$ is the number of memristors in the feedback loop ($N=2$ in this case), $R_\textrm{in}$ is the resistance placed at the input of the inverting pin of the OpAmp (see \cref{fig:concept}), and $V_\textrm{in}$ is the input voltage of the prototype. The analog switches are set by the GPIOs of the LOTUS board to establish connections between the memristors and the APMUs allowing their programming. All memristors except one (i.e., $N-1$ memristors) are programmed to $R_\textrm{trg}$ using a resistance tuning read-write-verify algorithm from Alibart et al. \cite{Alibart2012}. At each step of the algorithm, a \SI{200}{\nano\second} write pulse is applied, with its polarity either increasing the resistance (negative amplitude) or decreasing the  resistance (positive amplitude). A \SI{10}{\micro\second}/$V_\textrm{r}$ read pulse is then applied to measure the new memristor resistance. If $R_\textrm{trg}$ is not reached, the next write pulse amplitude is linearly increased by $ s_V=\SI{10}{\milli\volt}$. Once the target resistance is reached within a tolerance of 1\%, 10 read pulses are applied to the memristor to ensure the stability of the programmed state. The read pulse amplitude, $V_\textrm{r}$, is chosen to be the difference $V_\textrm{out}^\textrm{trg}-V_\textrm{in}$, as this is the voltage that will be applied to the memristors in the feedback loop of the TIA. If a typical $V_r$ value of \SI{0.2}{\volt} is used, it will be impossible to accurately tune the memristor resistances to achieve a given $V_\textrm{out}^\textrm{trg}$ value due to the I-V nonlinearities of memristors \cite{Wang2021}. The final memristor is programmed to balance the error accumulated during the programming of the other ($N-1$) memristors. This balancing resistance state is given by:
\begin{equation}
    R_\textrm{bal} = \left(\frac{N}{R_\textrm{trg}}-\sum_i^{N-1}G^\textrm{m}_i\right)^{-1}
\end{equation}
where $\frac{N}{R_\textrm{trg}}$ is the target feedback conductance and $\sum_i^{N-1}G^\textrm{m}_i$ is the sum of the conductance programmed on the $N-1$ memristors. The same resistance tuning algorithm is used with a smaller tolerance (0.5\%) to improve the final accuracy of the programmable DC source. Finally, the analog switches are set by the GPIOs to connect the memristors to the feedback loop of the TIA. To evaluate the performance of the circuit, a constant $V_\textrm{in}$ is applied and $V_\textrm{out}$ is measured. The $V_\textrm{in}$ is fixed for all programmed output voltages.\\

For the cryogenic measurements, the cryogenic DC source PCB (see \cref{fig:Proto_exp_setup}b) is placed in the \SI{1}{\kelvin}-pot of an ICE Oxford DRY ICE cryostat. The LOTUS board and the analog switches are placed outside the cryostat in an electronic rack. The memristors and feedback loop lines are connected to 5 BeCu RF lines while the supply voltages, $V_\textrm{in}$ and $V_\textrm{out}$, are connected to 4 superconducting DC lines with a cut-off frequency around \SI{10}{\mega\hertz}. 

%% file: 05_Measurement.tex
\begin{figure*}
\center{\includegraphics[width=0.9\linewidth]{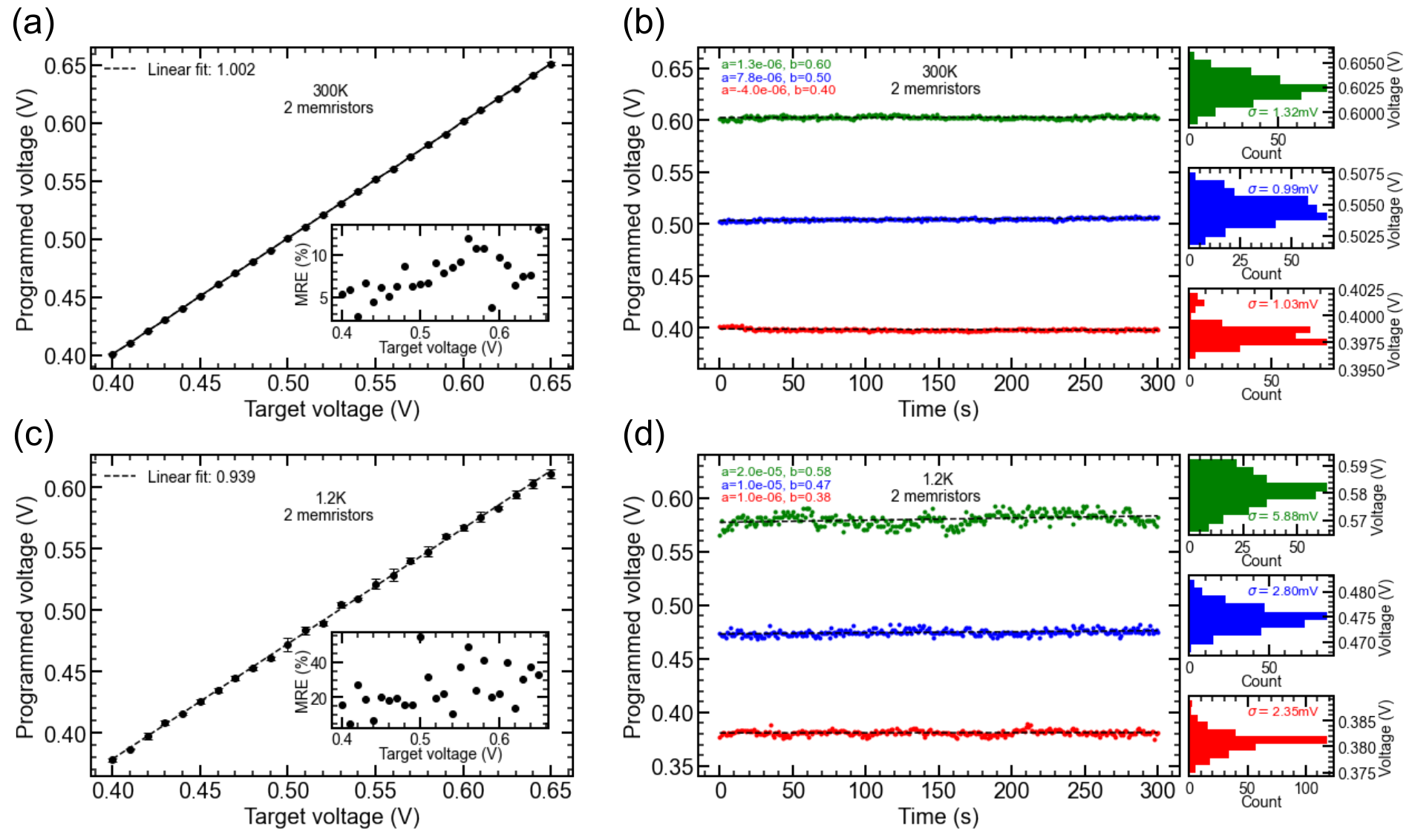}}
  \caption{\label{fig:dc_source_cara}\textbf{Electrical characteristics of the memristor-based DC source prototype.} \textbf{a.} A \SI{250}{\milli\volt} voltage sweep between \SI{0.4}{\volt} and \SI{0.65}{\volt} with a resolution of \SI{10}{\milli\volt} at \SI{300}{\kelvin}. The experiment was performed with 2 memristors connected in parallel in the feedback loop of the TIA circuit, with $V_\textrm{in}=\SI{0.25}{\volt}$ and $V_\textrm{dd}=-V_\textrm{ss}=\SI{2.7}{\volt}$. The voltage sweep is performed 10 times, with each point representing the mean voltage programmed while the error bars show the programming error. The mean resolution error (MRE) is shown in the inset. \textbf{b.} Stability of three intermediary programmed voltages at \SI{300}{\kelvin}. Each programmed voltage is fitted by linear regression ($aV+b$) to verify the drift of the programmed voltage. The three insets depict the variability of each stable programmed voltage. \textbf{c.}  Ten \SI{250}{\milli\volt} voltage sweeps between \SI{0.4}{\volt} and \SI{0.65}{\volt} with a resolution of \SI{10}{\milli\volt} at \SI{1.2}{\kelvin}. The experiment was performed with 2 memristors connected in parallel in the feedback loop of the TIA circuit, with $V_\textrm{in}=\SI{75}{\milli\volt}$ and $V_\textrm{dd}=-V_\textrm{ss}=\SI{3.0}{\volt}$. The inset shows the MRE for this measurement. \textbf{d.} Stability of three programmed voltages at \SI{1.2}{\kelvin}. The three insets depict the variability of each stable programmed voltage.}

\end{figure*}

\section{Experimental results}
In order to validate the concept of the cryogenic memristor-based DC source, the prototype has to demonstrate output voltage tunability with limited error (10\% of the voltage resolution) in accordance with QD biasing requirements i.e., a \SI{0.25}-\SI{1}{\volt} output range. Hence, we performed DC sweep measurements at \SI{300}{\kelvin} as a performance benchmark. The output voltage is swept between \SI{0.4}{\volt} and \SI{0.65}{\volt} with a resolution of  \SI{10}{\milli\volt}. This is achieved by running the algorithm described in \cref{fig:Proto_exp_setup}c with $V_\textrm{in}=\SI{0.25}{\volt}$ and $N=2$. This voltage sweep corresponds to tuning each memristor resistance between \SI{9.6}{\kilo\ohm} and \SI{15.6}{\kilo\ohm}. We perform 10 DC sweep measurements to assess the programming variability of the memristor-based DC source. The mean programmed voltages are reported in \cref{fig:dc_source_cara}a, demonstrating the tunability of the DC source over the \SI{0.4}-\SI{0.65}{\volt} range. The DC offset ($\approx\SI{8}{\milli\volt}$) introduced by the AD8605 operational amplifier is subtracted from the programmed voltages to mitigate systematic error. The mean programmed voltages are fitted by a linear function to verify the linearity of the output voltage. Additionally, the fitted slope ($a_f=1.002$) allows for the quantification of the memristor feedback resistance programming accuracy. As the fitted slope is close to one, the gain of the TIA is programmed accurately. Moreover, the mean resolution error (MRE) is computed for each programmed output voltage. It is is given by:
\begin{equation}
    \textrm{MRE}(V_\textrm{out})=100\times\frac{\textrm{std}(V_\textrm{out})}{a_\textrm{f}\delta V}
\end{equation}
where $\textrm{std}(V_\textrm{out})$ is the standard deviation of a given output voltage $V_\textrm{out}$ over 10 sweep measurements, $a_\textrm{f}$ is the fitted slope and $\delta V$ is the intended voltage resolution for the sweep. The MRE for the \SI{300}{\kelvin} DC sweeps is shown in the inset of \cref{fig:dc_source_cara}a with an average MRE below 10\% i.e., the programmed output voltage is tuned with less than a \SI{1}{\milli\volt} error.\\

While the DC sweep measurements validate the tunability of the memristor-based DC source prototype, it is necessary to verify the stability of the programmed voltages. This can be done by measuring an arbitrary programmed output voltage over several seconds e.g., \SI{300}{\second} for the stability measurement reported in \cref{fig:dc_source_cara}b. The time traces of the programmed voltages are fitted by a linear function ($V(t)=at+b$) to assess the stability of the output. The fitted slope factor is on the order of $5\times10^{-5}\SI{}{\volt\per\second}$ which indicates a \SI{5}{\micro\volt} drift every second. This drift is compatible with QD biasing as the drift timescale is very large with respect to spin qubits coherence time \cite{Veldhorst2014}. These stability measurements also allows for the evaluation of the voltage noise amplitude exhibited by the memristor-based DC source prototype.
The three insets of \cref{fig:dc_source_cara}b show a histogram of each programmed output voltage, indicating that the voltage noise amplitude follows a Gaussian distribution with a standard deviation around \SI{1}{\milli\volt} for the three tested output voltages.\\

The same set of measurements are performed at \SI{1.2}{\kelvin} to verify the cryogenic compatibility of the memristor-based DC source prototype. Under these conditions, the TiO$_\textrm{x}$ memritors used in the feedback loop need to be reformed at cryogenic temperatures to remove the metal-insulator transition hindering their analog programmability as suggested by Ref.\cite{MounyAPL2023}. The 10 DC sweeps measured at \SI{1.2}{\kelvin} are performed with a different input voltage ($V_\textrm{in}=\SI{75}{\milli\volt}$) due to the increased resistance of the cryo-reformed TiO$_\textrm{x}$ memristors at cryogenic temperatures \cite{MounyAPL2023}. The same DC sweep is attempted at \SI{1.2}{\kelvin} i.e., a \SI{0.4}{\volt}-\SI{0.65}{\volt} output range with a \SI{10}{\milli\volt} resolution. Performing this DC sweep requires to program the two memristors between \SI{32}{\kilo\ohm} and \SI{52}{\kilo\ohm}.  The voltage supply of the AD8605 OpAmp is increased to $\pm\SI{3.0}{\volt}$ to compensate for its gain loss at cryogenic temperatures (See \cref{fig:AD8605_carac}c). This is not sufficient to achieve the targeted gain at each step of the sweep as the fitted slope ($a_f=0.939$) suggests. Moreover, the inset of \cref{fig:dc_source_cara}c suggests that the higher resistances of the memristors at cryogenic temperatures limits the performance of the memristor-based DC source prototype as the average MRE is approximately 2.5 times larger than the MRE at \SI{300}{\kelvin}. Additionally, the current consumption of the AD8605 OpAmp was measured during the DC voltage sweeps and ranges from \SI{1.43}{\milli\ampere} to \SI{1.84}{\milli\ampere}. This yields to a power consumption of around \SI{10}{\milli\watt} for the OpAmp, while the memristor resistive feedback dissipates between \SI{6}{\micro\watt} and \SI{12}{\micro\watt}. Assuming that most of the power consummed by the OpAmp is dissipated, the AD8605 introduces a bottleneck in power dissipation if the prototype was to be scaled up. Nonetheless, the \cref{fig:dc_source_cara}c demonstrates the viability of the memristor-based DC source concept.\\

The stability of the programmed DC voltage was also investigated at \SI{1.2}{\kelvin} (see \cref{fig:dc_source_cara}d). The memristor-based DC source shows a lower voltage drift at cryogenic temperature as the slope factor $a$ is up to an order of magnitude smaller than the one fitted at \SI{300}{\kelvin}. This increases the retention of the DC source prototype to a \SI{1}{\milli\volt} drift every \SI{1000}{\second}, which is a very large retention time when compared to the coherence time of spin qubits ($\approx\SI{10}{\milli\second}$\cite{Veldhorst2014}). This validates the edge of the memristor-based DC source concept over switched-capacitors circuits which need to be refreshed every tens of microseconds \cite{Geck2019}. However, the programmed output voltages exhibit a larger voltage noise amplitude. This is mainly due to the larger resistance of the TiO$_\textrm{x}$ memristors at \SI{1.2}{\kelvin} as suggested previously \cite{MounyAPL2023}. The read variability introduced by memristors increases linearly with their resistance \cite{Marcotte2023}. This is  validated by the \SI{300}{\kelvin} measurements which exhibit a voltage noise up to five times smaller with smaller memristor resistances. Electronic noise is the main concern when interfacing with quantum dots as an output voltage noise too important could lead to decoherence of the qubit \cite{Kuhlman2013, Connors2019}. This output voltage noise could be mitigated by using a low-pass filter which would not prevent the operation of the DC source as its role is to apply a DC bias to quantum dots. However, it is to be noted that such filtering would add an additional thermal load.\\

While the memristor-based DC source prototype shows an insufficient voltage resolution to achieve quantum dot biasing, our previous study suggests that increasing linearly the number of memristors placed in feedback loop allows to reduce the voltage resolution exponentially \cite{Mouny2023TED}. Namely using 8 memristors would allow to reach a voltage resolution of $\approx\SI{100}{\micro\volt}$. This could be achieved by monolithically integrating the memristors in the feedback loop of an integrated CMOS OpAmp \cite{mesoudy2021CMOS}.

%% file: 06_Scaling.tex
\begin{figure*}
\center{\includegraphics[width=0.9\linewidth]{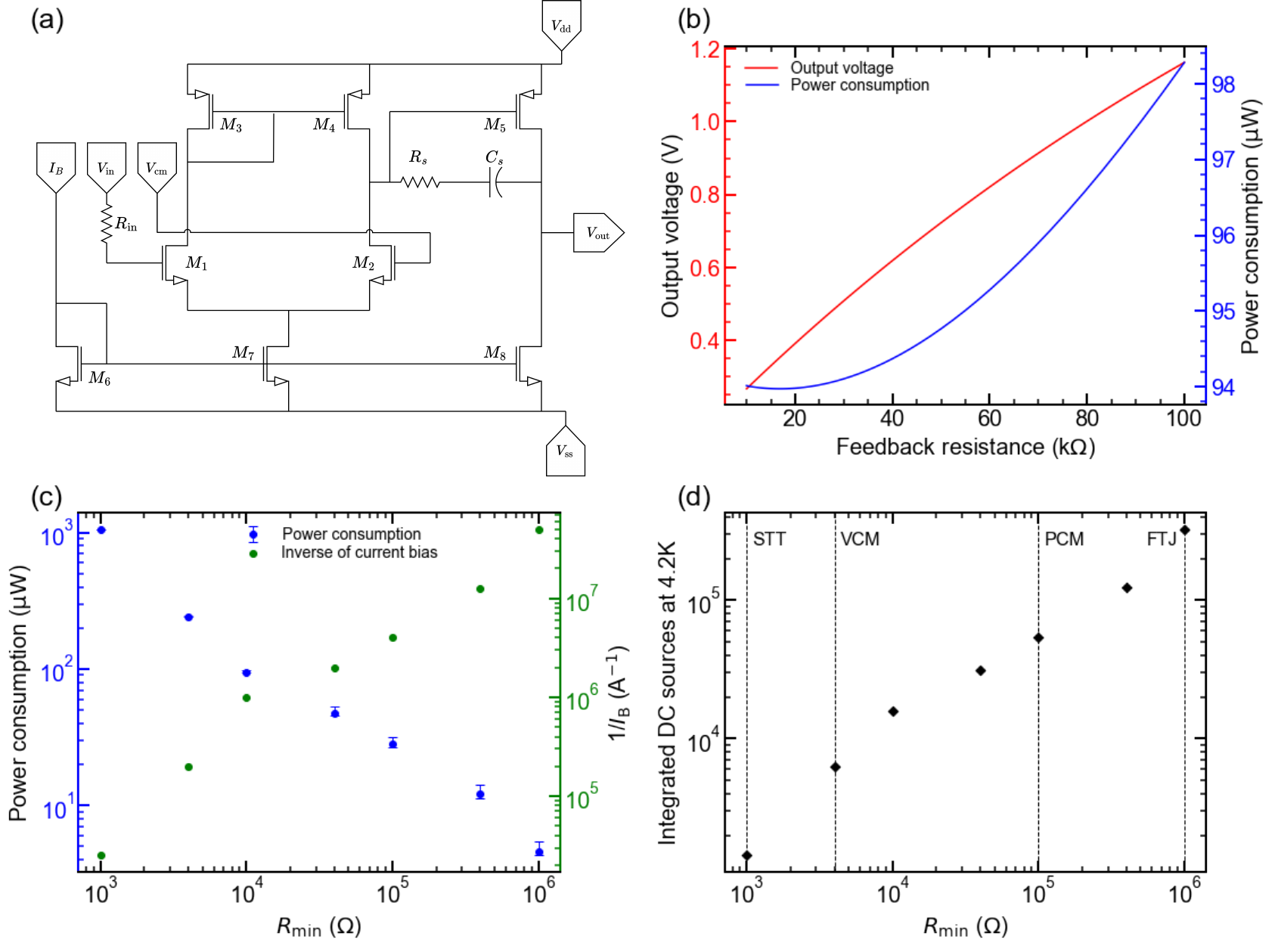}}  \caption{\label{fig:scaling} \textbf{eNVM-based DC source scaling simulations.} \textbf{a.} The operational amplifier schematic based on a two-stage Miller topology. One or multiple eNVM can be placed between the V$_\textrm{in}$ and V$_{\textrm{out}}$ nodes to enable the tunability of the DC source. A single fixed resistor R$_{\textrm{in}}$ is placed after the V$_\textrm{in}$ node to allow transimpedance amplification. \textbf{b.} DC characteristics of the operational amplifier for a feedback resistance ranging from \SI{10}{\kilo\ohm} to \SI{100}{\kilo\ohm}. The circuit parameters used for this simulation are reported in \cref{tab:params}. \textbf{c} Feedback resistance scaling simulation. Larger resistances results in a decrease of the bias current I$_\textrm{B}$, leading to a power consumption decrease. \textbf{d} Maximum number of eNVM-based DC sources integrable at the \SI{4.2}{\kelvin} stage of a Bluefors XLD dilution fridge i.e., a \SI{1.5}{\watt} cooling power. The dashed black lines show the minimum resistance of different eNVM technologies.}
  \end{figure*}

\section{Scaling discussions}
The previous study performed with discrete components and a limited number of memristors serves as a proof of concept for this novel cryogenic DC source. Making this solution scalable now requires the design of a fully integrated CMOS-memristor circuit. Moving from discrete electronic components to integrated electronics will drastically reduce both the power consumption and footprint of the operational amplifier used in the memristor-based DC source. The scaling of this DC source is investigated using electronic computer-aided design and circuit simulations to estimate the DC source footprint, power consumption, and electrical characteristics. Firstly, an integrated TIA in \SI{65}{\nano\metre} TSMC CMOS technology is designed (see \cref{fig:scaling}a). This TIA is based on a two-stage Miller operational amplifier \cite{Palmisano2001} with a memristor resistive feedback, a concept that can be extended to all emerging non-volatile memories (eNVM). For scaling purposes, the memristors should be fabricated in the back end of line of the TIA CMOS chip using CMOS-compatible fabrication processes. The initial TIA is designed to be used with a single CMOS-compatible TiO$_\textrm{x}$ memristor from Ref. \cite{MounyAPL2023}, more precisely called valence change material memory (VCM), as the resistive feedback. These VCMs demonstrate analog programmability down to \SI{4.2}{\kelvin} between \SI{10}{\kilo\ohm} to \SI{100}{\kilo\ohm} \cite{MounyAPL2023}. The design parameters are reported in \cref{tab:params}.\\

\begin{table}[b]
\centering
\caption{\label{tab:params} Parameters for the OpAmp design used in \cref{fig:scaling}(b).  All transistor length are set to \SI{65}{\nano\metre}. For all simulations in \cref{fig:scaling}(c) and (d), R$_\textrm{in}$ is equal to R$_\textrm{min}$/4.}
\begin{tabular}{p{2cm}p{2cm}p{2cm}}
\hline
\hline
Parameter&Value&Unit\\
\hline
I$_\textrm{B}$ & 1 &\SI{}{\micro\ampere}\\
V$_\textrm{in}$, V$_\textrm{cm}$ & 100&\SI{}{\milli\volt}\\
V$_\textrm{dd}$, V$_\textrm{ss}$ & $\pm$3   & \SI{}{\volt} \\
M$_1$, M$_2$  & 1.5 &\SI{}{\micro\metre}\\
M$_3$, M$_4$  & 1 &\SI{}{\micro\metre} \\
M$_5$  & 10 &\SI{}{\micro\metre} \\
M$_6$  & 380 &\SI{}{\nano\metre} \\
M$_7$  & 760 &\SI{}{\nano\metre} \\
M$_8$  & 120 &\SI{}{\nano\metre} \\
R$_\textrm{s}$  & 10 &\SI{}{\kilo\ohm} \\
C$_\textrm{s}$  &  3.5&\SI{}{\pico\farad}\\
R$_\textrm{in}$  & 2.5&\SI{}{\kilo\ohm} \\
\hline
\hline
\end{tabular}
\end{table}

The first stage of the TIA is composed of a stabilized differential pair biased with $2\textrm{I}_\textrm{B}$ by the $M_6-M_7$ current mirror. The second stage is a simple source follower biased by $\frac{1}{3}\textrm{I}_\textrm{B}$. From the room temperature DC simulations, a biasing current $\textrm{I}_\textrm{B}$ of \SI{1}{\micro\ampere} is needed to achieve a \SI{1}{\volt} output range which is required for quantum dot auto-tuning. This setup yields to a sub-\SI{100}{\micro\watt} power consumption for a single eNVM-based DC source (See \cref{fig:scaling}(b)) i.e., a 80 times smaller power consumption than the discrete prototype experimentally tested. Using the same design, the power consumption can be reduced by increasing the feedback resistance while decreasing the bias current $\textrm{I}_\textrm{B}$. As depicted in \cref{fig:scaling}(c), this approach allows to reduce the power consumption by an additional factor of 20 by lowering $\textrm{I}_\textrm{B}$ to \SI{20}{\nano\ampere}. However for $R_\textrm{min}\geq$\SI{50}{\kilo\ohm}, the width of $M_5$ is increased to \SI{40}{\micro\metre} to maintain the \SI{0.2}-\SI{1.2}{\volt} output range. This increases the footprint of the TIA by a factor of approximately 4 which could limit the integration density. This level of power consumption enables a significant scaling up of the number of eNVM-DC sources integrated at the \SI{4.2}{\kelvin} stage of a dilution fridge close to 'hot' spin qubits \cite{Camenzind2022}. Utilizing the same TiO$_\textrm{x}$ VCM memory \cite{MounyAPL2023} would facilitate the integration of up to 16,000 VCM-based DC sources (see \cref{fig:scaling}(d)), based on a \SI{1.5}{\watt} cooling power and considering that all consumed power needs to be dissipated. However, using alternate eNVMs, such as ferroelectric tunnel junctions (FTJ) which exhibit resistances above \SI{1}{\mega\ohm} \cite{Boyn2014} and working at cryogenic temperatures \cite{Hur2022}, will allow to integrate up to 300,000 DC sources at \SI{4.2}{\kelvin}. This is mainly due to the larger resistance of FTJs enabling to lower the current bias of the amplifier and thus its power consumption.  Additionally, this TIA topology has been demonstrated down to \SI{4.2}{\kelvin} in a smaller \SI{28}{\nano\metre} fully depleted silicon on insulator (FDSOI) technological node \cite{cryo_ampli_le_guevel_2020}. This TIA exhibited a power consumption of \SI{1}{\micro\watt} for a feedback resistance in the same order of FTJ resistances. Therefore designing a custom TIA in \SI{28}{\nano\metre} FDSOI node with higher eNVM resistances would allow to reduce the power consumption down to $\approx\SI{1}{\micro\watt}$ per DC source i.e., by almost 4 order of magnitude compared to the experimental prototype presented. This would enable the control of nearly to one million quantum dots, assuming two gates require biasing by quantum dots \cite{Geck2019}.

%% file: 07_Conclusion.tex
\section{Conclusion}

In conclusion, we validate the viability of a memristor-based cryogenic programmable DC source for scalable \textit{in-situ} quantum-dot control. The cryogenic compatibility of the commercial operational amplifier AD8605 is tested down to \SI{1.2}{\kelvin}. At cryogenic temperatures the AD8605 exhibit a decrease in voltage gain and an increase in current consumption. 
Additionally, we demonstrate that this simple control approach which includes an OpAmp and memristor devices allows to perform \SI{0.25}{\volt}-DC sweeps with a \SI{10}{\milli\volt} voltage resolution with only 2 memristors at \SI{1.2}{\kelvin}. The memristor-based DC source prototype shows an output voltage retention time well above the coherence time of spin qubits which is the main advantage of this concept over switched-capacitor circuits. To fulfill the baseline requirements for quantum dot biasing (i.e., a \SI{1}{\milli\volt}-resolution over a \SI{1}{\volt}-range), monolithically co-integrate memristors with advanced CMOS circuitry (OpAmp and analog switches)  can be utilized to enable low power dissipation down to a few microwatts per DC source within a small footprint, closing the power consumption gap with switched-capacitors biasing circuit . By using alternate emerging non-volatile memory technologies like FTJs, scaling up this concept encompass a few hundred of thousands of eNVM-based DC source for in situ quantum dot biasing paving the way for large-scale quantum computing applications.\\

%% file: 08_Others.tex
\noindent\textbf{Data availability}
The data underlying the results presented in this paper are not
publicly available at this time but may be obtained from the authors
upon reasonable request.\\
\linebreak
\textbf{Acknowledgements}
This work was supported by Natural Sciences and Engineering Research Council of Canada (NSERC). This research was undertaken thanks in part to funding from the Canada First Research Excellence Fund. LN2 is French-Canadian joint International Research Laboratory (IRL-3463) funded and co-operated by CNRS, Université de Sherbrooke, Université de Grenoble Alpes (UGA), École Centrale Lyon (ECL) and INSA Lyon. It is supported by the Fonds de Recherche du Québec Nature et Technologie (FRQNT). We would like to acknowledge CMC Microsystems for the provision of products and services that facilitated this research, including CAD tools and 65-nm CMOS technology PDK. We would like to thank Christian Lupien, Simon Fortier and the Institut Quantique for their support with the electrical characterisation at cryogenic temperatures.
\\

\noindent\textbf{Author contributions}\\
\textbf{Pierre-Antoine Mouny:} Data curation; Investigation; Methodology; Resources; Software; Visualization; Writing – original draft; Writing – review \& editing.
\textbf{Raphaël Dawant:} Resources; Writing – review \& editing.
\textbf{Patrick Dufour:} Resources; Software; Writing – review \& editing.
\textbf{Matthieu Valdenaire:} Software; Writing – review \& editing.
\textbf{Serge Ecoffey:} Project administration; Supervision; Writing – review \& editing.
\textbf{Michel Pioro-Ladrière:} Funding acquisition; Project administration; Writing – review \& editing.
\textbf{Yann Beilliard:} Conceptualization; Project administration; Supervision; Writing – review \& editing.
\textbf{Dominique Drouin:} Conceptualization; Funding acquisition; Project administration; Supervision; Writing – review \& editing.
\linebreak
\textbf{Competing interests}
The authors declare no competing interests.

%% file: main.bbl
\begin{thebibliography}{10}

\bibitem{AspuruGuzik2005}
Alan Aspuru-Guzik, Anthony~D. Dutoi, Peter~J. Love, and Martin Head-Gordon.
\newblock Simulated quantum computation of molecular energies.
\newblock {\em Science}, 309(5741):1704--1707, September 2005.

\bibitem{Ors2019}
Rom{\'{a}}n Or{\'{u}}s, Samuel Mugel, and Enrique Lizaso.
\newblock Quantum computing for finance: Overview and prospects.
\newblock {\em Reviews in Physics}, 4:100028, November 2019.

\bibitem{berger2021quantum}
Casey Berger, Agustin~Di Paolo, Tracey Forrest, Stuart Hadfield, Nicolas Sawaya, Michal Stechly, and Karl Thibault.
\newblock Quantum technologies for climate change: Preliminary assessment, 2021.

\bibitem{Nakamura1999}
Y.~Nakamura, Yu.~A. Pashkin, and J.~S. Tsai.
\newblock Coherent control of macroscopic quantum states in a single-{C}ooper-pair box.
\newblock {\em Nature}, 398(6730):786–788, April 1999.

\bibitem{Arute2019}
Frank Arute, Kunal Arya, Ryan Babbush, Dave Bacon, Joseph~C. Bardin, Rami Barends, Rupak Biswas, Sergio Boixo, Fernando G. S.~L. Brandao, David~A. Buell, Brian Burkett, Yu~Chen, Zijun Chen, Ben Chiaro, Roberto Collins, William Courtney, Andrew Dunsworth, Edward Farhi, Brooks Foxen, Austin Fowler, Craig Gidney, Marissa Giustina, Rob Graff, Keith Guerin, Steve Habegger, Matthew~P. Harrigan, Michael~J. Hartmann, Alan Ho, Markus Hoffmann, Trent Huang, Travis~S. Humble, Sergei~V. Isakov, Evan Jeffrey, Zhang Jiang, Dvir Kafri, Kostyantyn Kechedzhi, Julian Kelly, Paul~V. Klimov, Sergey Knysh, Alexander Korotkov, Fedor Kostritsa, David Landhuis, Mike Lindmark, Erik Lucero, Dmitry Lyakh, Salvatore Mandr{\`{a}}, Jarrod~R. McClean, Matthew McEwen, Anthony Megrant, Xiao Mi, Kristel Michielsen, Masoud Mohseni, Josh Mutus, Ofer Naaman, Matthew Neeley, Charles Neill, Murphy~Yuezhen Niu, Eric Ostby, Andre Petukhov, John~C. Platt, Chris Quintana, Eleanor~G. Rieffel, Pedram Roushan, Nicholas~C. Rubin, Daniel Sank, Kevin~J.
  Satzinger, Vadim Smelyanskiy, Kevin~J. Sung, Matthew~D. Trevithick, Amit Vainsencher, Benjamin Villalonga, Theodore White, Z.~Jamie Yao, Ping Yeh, Adam Zalcman, Hartmut Neven, and John~M. Martinis.
\newblock Quantum supremacy using a programmable superconducting processor.
\newblock {\em Nature}, 574(7779):505--510, October 2019.

\bibitem{Cirac1995}
J.~I. Cirac and P.~Zoller.
\newblock Quantum computations with cold trapped ions.
\newblock {\em Phys. Rev. Lett.}, 74:4091--4094, May 1995.

\bibitem{Leibfried2003}
D.~Leibfried, R.~Blatt, C.~Monroe, and D.~Wineland.
\newblock Quantum dynamics of single trapped ions.
\newblock {\em Rev. Mod. Phys.}, 75:281--324, Mar 2003.

\bibitem{Bedecarrats2021}
T.~Bédécarrats, B.~Cardoso Paz, B.~Martinez Diaz, H.~Niebojewski, B.~Bertrand, N.~Rambal, C.~Comboroure, A.~Sarrazin, F.~Boulard, E.~Guyez, J.-M. Hartmann, Y.~Morand, A.~Magalhaes-Lucas, E.~Nowak, E.~Catapano, M.~Cassé, M.~Urdampilleta, Y.-M. Niquet, F.~Gaillard, S.~De~Franceschi, T.~Meunier, and M.~Vinet.
\newblock A new {FDSOI} spin qubit platform with 40nm effective control pitch.
\newblock In {\em 2021 IEEE International Electron Devices Meeting (IEDM)}, pages 1--4, 2021.

\bibitem{Veldhorst2014}
M.~Veldhorst, J.~C.~C. Hwang, C.~H. Yang, A.~W. Leenstra, B.~de~Ronde, J.~P. Dehollain, J.~T. Muhonen, F.~E. Hudson, K.~M. Itoh, A.~Morello, and A.~S. Dzurak.
\newblock An addressable quantum dot qubit with fault-tolerant control-fidelity.
\newblock {\em Nature Nanotechnology}, 9(12):981--985, October 2014.

\bibitem{Petit2020}
L.~Petit, H.~G.~J. Eenink, M.~Russ, W.~I.~L. Lawrie, N.~W. Hendrickx, S.~G.~J. Philips, J.~S. Clarke, L.~M.~K. Vandersypen, and M.~Veldhorst.
\newblock Universal quantum logic in hot silicon qubits.
\newblock {\em Nature}, 580(7803):355--359, April 2020.

\bibitem{Yang2020}
C.~H. Yang, R.~C.~C. Leon, J.~C.~C. Hwang, A.~Saraiva, T.~Tanttu, W.~Huang, J.~Camirand Lemyre, K.~W. Chan, K.~Y. Tan, F.~E. Hudson, K.~M. Itoh, A.~Morello, M.~Pioro-Ladri{\`{e}}re, A.~Laucht, and A.~S. Dzurak.
\newblock Operation of a silicon quantum processor unit cell above one kelvin.
\newblock {\em Nature}, 580(7803):350--354, April 2020.

\bibitem{Camenzind2022}
Leon~C. Camenzind, Simon Geyer, Andreas Fuhrer, Richard~J. Warburton, Dominik~M. Zumb\"{u}hl, and Andreas~V. Kuhlmann.
\newblock A hole spin qubit in a fin field-effect transistor above 4{\hspace{0.167em}}kelvin.
\newblock {\em Nature Electronics}, 5(3):178--183, March 2022.

\bibitem{Stuyck2021}
N.~I.~Dumoulin Stuyck, R.~Li, C.~Godfrin, A.~Elsayed, S.~Kubicek, J.~Jussot, B.~T. Chan, F.~A. Mohiyaddin, M.~Shehata, G.~Simion, Y.~Canvel, L.~Goux, M.~Heyns, B.~Govoreanu, and I.~P. Radu.
\newblock Uniform spin qubit devices with tunable coupling in an all-silicon 300 mm integrated process.
\newblock In {\em 2021 Symposium on VLSI Circuits}, pages 1--2, 2021.

\bibitem{Zwerver_2022}
A.~M.~J. Zwerver, T.~Krähenmann, T.~F. Watson, L.~Lampert, H.~C. George, R.~Pillarisetty, S.~A. Bojarski, P.~Amin, S.~V. Amitonov, J.~M. Boter, R.~Caudillo, D.~Correas-Serrano, J.~P. Dehollain, G.~Droulers, E.~M. Henry, R.~Kotlyar, M.~Lodari, F.~Lüthi, D.~J. Michalak, B.~K. Mueller, S.~Neyens, J.~Roberts, N.~Samkharadze, G.~Zheng, O.~K. Zietz, G.~Scappucci, M.~Veldhorst, L.~M.~K. Vandersypen, and J.~S. Clarke.
\newblock Qubits made by advanced semiconductor manufacturing.
\newblock {\em Nature Electronics}, 5(3):184--190, mar 2022.

\bibitem{XuSwitchedCapa2020}
Y.~Xu, F.~K. Unseld, A.~Corna, A.~M.~J. Zwerver, A.~Sammak, D.~Brousse, N.~Samkharadze, S.~V. Amitonov, M.~Veldhorst, G.~Scappucci, R.~Ishihara, and L.~M.~K. Vandersypen.
\newblock On-chip integration of {Si/SiGe}-based quantum dots and switched-capacitor circuits.
\newblock {\em Applied Physics Letters}, 117(14):144002, 2020.

\bibitem{Fowler_2012}
Austin~G. Fowler, Matteo Mariantoni, John~M. Martinis, and Andrew~N. Cleland.
\newblock Surface codes: Towards practical large-scale quantum computation.
\newblock {\em Physical Review A}, 86(3), 2012.

\bibitem{Xue2022}
Xiao Xue, Maximilian Russ, Nodar Samkharadze, Brennan Undseth, Amir Sammak, Giordano Scappucci, and Lieven M.~K. Vandersypen.
\newblock Quantum logic with spin qubits crossing the surface code threshold.
\newblock {\em Nature}, 601(7893):343–347, January 2022.

\bibitem{Lawrie2020}
W.~I.~L. Lawrie, H.~G.~J. Eenink, N.~W. Hendrickx, J.~M. Boter, L.~Petit, S.~V. Amitonov, M.~Lodari, B.~Paquelet~Wuetz, C.~Volk, S.~G.~J. Philips, G.~Droulers, N.~Kalhor, F.~van Riggelen, D.~Brousse, A.~Sammak, L.~M.~K. Vandersypen, G.~Scappucci, and M.~Veldhorst.
\newblock Quantum dot arrays in silicon and germanium.
\newblock {\em Applied Physics Letters}, 116(8), February 2020.

\bibitem{Borsoi2023}
Francesco Borsoi, Nico~W. Hendrickx, Valentin John, Marcel Meyer, Sayr Motz, Floor van Riggelen, Amir Sammak, Sander~L. de~Snoo, Giordano Scappucci, and Menno Veldhorst.
\newblock Shared control of a 16~semiconductor quantum dot crossbar array.
\newblock {\em Nature Nanotechnology}, 19(1):21–27, August 2023.

\bibitem{Ansaloni2020}
Fabio Ansaloni, Anasua Chatterjee, Heorhii Bohuslavskyi, Benoit Bertrand, Louis Hutin, Maud Vinet, and Ferdinand Kuemmeth.
\newblock Single-electron operations in a foundry-fabricated array of quantum dots.
\newblock {\em Nature Communications}, 11(1), December 2020.

\bibitem{Geyer2021}
Simon Geyer, Leon~C. Camenzind, Lukas Czornomaz, Veeresh Deshpande, Andreas Fuhrer, Richard~J. Warburton, Dominik~M. Zumb\"{u}hl, and Andreas~V. Kuhlmann.
\newblock Self-aligned gates for scalable silicon quantum computing.
\newblock {\em Applied Physics Letters}, 118(10), March 2021.

\bibitem{Charbon2016}
E.~Charbon, F.~Sebastiano, A.~Vladimirescu, H.~Homulle, S.~Visser, L.~Song, and R~M. Incandela.
\newblock Cryo-{CMOS} for quantum computing.
\newblock In {\em 2016 IEEE International Electron Devices Meeting (IEDM)}, pages 13.5.1--13.5.4, 2016.

\bibitem{Patra2018}
Bishnu Patra, Rosario~M. Incandela, Jeroen P.~G. van Dijk, Harald A.~R. Homulle, Lin Song, Mina Shahmohammadi, Robert~Bogdan Staszewski, Andrei Vladimirescu, Masoud Babaie, Fabio Sebastiano, and Edoardo Charbon.
\newblock Cryo-{CMOS} circuits and systems for quantum computing applications.
\newblock {\em IEEE Journal of Solid-State Circuits}, 53(1):309--321, 2018.

\bibitem{Xue2021}
Xiao Xue, Bishnu Patra, Jeroen P.~G. van Dijk, Nodar Samkharadze, Sushil Subramanian, Andrea Corna, Brian~Paquelet Wuetz, Charles Jeon, Farhana Sheikh, Esdras Juarez-Hernandez, Brando~Perez Esparza, Huzaifa Rampurawala, Brent Carlton, Surej Ravikumar, Carlos Nieva, Sungwon Kim, Hyung-Jin Lee, Amir Sammak, Giordano Scappucci, Menno Veldhorst, Fabio Sebastiano, Masoud Babaie, Stefano Pellerano, Edoardo Charbon, and Lieven M.~K. Vandersypen.
\newblock {CMOS}-based cryogenic control of silicon quantum circuits.
\newblock {\em Nature}, 593(7858):205--210, May 2021.

\bibitem{Pauka2021}
S.~J. Pauka, K.~Das, R.~Kalra, A.~Moini, Y.~Yang, M.~Trainer, A.~Bousquet, C.~Cantaloube, N.~Dick, G.~C. Gardner, M.~J. Manfra, and D.~J. Reilly.
\newblock A cryogenic {CMOS} chip for generating control signals for multiple qubits.
\newblock {\em Nature Electronics}, 4(1):64--70, January 2021.

\bibitem{Geck2019}
L.~{Geck}, A.~{Kruth}, H.~{Bluhm}, S.~{van Waasen}, and S.~{Heinen}.
\newblock Control electronics for semiconductor spin qubits.
\newblock {\em Quantum Science and Technology}, 5(1), 2019.

\bibitem{Prabowo2021}
Bagas Prabowo, Guoji Zheng, Mohammadreza Mehrpoo, Bishnu Patra, Patrick Harvey-Collard, Jurgen Dijkema, Amir Sammak, Giordano Scappucci, Edoardo Charbon, Fabio Sebastiano, Lieven M.~K. Vandersypen, and Masoud Babaie.
\newblock 13.3 a 6-to-8{GHz} 0.17{mW}/qubit cryo-{CMOS} receiver for multiple spin qubit readout in 40nm cmos technology.
\newblock In {\em 2021 IEEE International Solid-State Circuits Conference (ISSCC)}, volume~64, pages 212--214, 2021.

\bibitem{ruffino2021integrated}
Andrea Ruffino, Tsung-Yeh Yang, John Michniewicz, Yatao Peng, Edoardo Charbon, and Miguel~Fernando Gonzalez-Zalba.
\newblock A cryo-{CMOS} chip that integrates silicon quantum dots and multiplexed dispersive readout electronics.
\newblock {\em Nature Electronics}, December 2021.

\bibitem{Mouny2023TED}
Pierre-Antoine Mouny, Yann Beilliard, Sébastien Graveline, Marc-Antoine Roux, Abdelouadoud~El Mesoudy, Raphaël Dawant, Pierre Gliech, Serge Ecoffey, Fabien Alibart, Michel Pioro-Ladrière, and Dominique Drouin.
\newblock Memristor-based cryogenic programmable {DC} sources for scalable in situ quantum-dot control.
\newblock {\em IEEE Transactions on Electron Devices}, 70(4):1989--1995, 2023.

\bibitem{Pickett2011}
Matthew~D. Pickett, Julien Borghetti, J.~Joshua Yang, Gilberto Medeiros-Ribeiro, and R.~Stanley Williams.
\newblock Coexistence of memristance and negative differential resistance in a nanoscale metal-oxide-metal system.
\newblock {\em Advanced Materials}, 23(15):1730--1733, 2011.

\bibitem{Beilliard2020}
Yann Beilliard, François Paquette, Frédéric Brousseau, Serge Ecoffey, Fabien Alibart, and Dominique Drouin.
\newblock Investigation of resistive switching and transport mechanisms of {Al$_\textrm{2}$O$_\textrm{3}$/TiO$_\textrm{2-x}$} memristors under cryogenic conditions (1.5 {K}).
\newblock {\em AIP Advances}, 10(2):025305, 2020.

\bibitem{MounyAPL2023}
Pierre-Antoine Mouny, Raphaël Dawant, Bastien Galaup, Serge Ecoffey, Michel Pioro-Ladri{\`{e}}re, Yann Beilliard, and Dominique Drouin.
\newblock Analog programming of {CMOS}-compatible {Al$_\textrm{2}$O$_\textrm{3}$}/{TiO$_\textrm{2-x}$} memristor at 4.2{\hspace{0.167em}}{K} after metal-insulator transition suppression by cryogenic reforming.
\newblock {\em Applied Physics Letters}, 123(16), October 2023.

\bibitem{Voronkovskii2019}
V~A Voronkovskii, V~S Aliev, A~K Gerasimova, and D~R Islamov.
\newblock Conduction mechanisms of {TaN}/{HfO$_\textrm{x}$}/{Ni} memristors.
\newblock {\em Materials Research Express}, 6(7):076411, April 2019.

\bibitem{Fang2015}
Runchen Fang, Wenhao Chen, Ligang Gao, Weijie Yu, and Shimeng Yu.
\newblock Low-temperature characteristics of {HfO$_\textrm{x}$}-based resistive random access memory.
\newblock {\em IEEE Electron Device Letters}, 36(6):567--569, 2015.

\bibitem{Lan2023}
Jun Lan, Zhixiong Li, Zhenjie Chen, Quanzhou Zhu, Wenhui Wang, Muhammad Zaheer, Jiqing Lu, Jinxuan Liang, Mei Shen, Peng Chen, Kai Chen, Guobiao Zhang, Zhongrui Wang, Feichi Zhou, Longyang Lin, and Yida Li.
\newblock Improved performance of {Hf$_\textrm{x}$Zn$_\textrm{y}$O}-based {RRAM} and its switching characteristics down to 4 {K} temperature.
\newblock {\em Advanced Electronic Materials}, 9(3), January 2023.

\bibitem{Zhang2014}
Ye~Zhang, Ning Deng, Huaqiang Wu, Zhiping Yu, Jinyu Zhang, and He~Qian.
\newblock Metallic to hopping conduction transition in {Ta$_\textrm{2}$O$_\textrm{5-x}$}/{TaO$_\textrm{y}$} resistive switching device.
\newblock {\em Applied Physics Letters}, 105(6):063508, 2014.

\bibitem{Proctor2015}
J.~E. Proctor, A.~W. Smith, T.~M. Jung, and S.~I. Woods.
\newblock High-gain cryogenic amplifier assembly employing a commercial {CMOS} operational amplifier.
\newblock {\em Review of Scientific Instruments}, 86(7):073102, July 2015.

\bibitem{Homulle2019}
Harald Homulle.
\newblock {\em Cryogenic electronics for the read-out of quantum processors}.
\newblock PhD thesis, TU Delft, 2019.

\bibitem{Patterson_2001}
R.~Patterson, A.~Hammoud, and S.~Gerber.
\newblock Performance of various types of resistors at low temperatures.
\newblock {\em NASA Glenn Res. Center, Cleveland, OH, USA}, GESS Rep. NAS3-00142, 2001.

\bibitem{AMT}
Advanced~Micro Testing.
\newblock {AMT} {LOTUS} control platfrom.

\bibitem{Dawant2024}
R.~Dawant, Matthieu Gaudreau, Marc-Antoine Roy, Pierre-Antoine Mouny, Matthieu Valdenaire, Pierre Gliech, Javier~Arias Zapata, Fabien Alibart, Dominique Drouin, and Serge Ecoffey.
\newblock Damascene versus subtractive line cmp process for resistive memory crossbars beol integration.
\newblock {\em Micro and Nano Engineering}, page 100251, 2024.

\bibitem{Alibart2012}
Fabien Alibart, Ligang Gao, Brian~D Hoskins, and Dmitri~B Strukov.
\newblock High precision tuning of state for memristive devices by adaptable variation-tolerant algorithm.
\newblock {\em Nanotechnology}, 23(7):075201, January 2012.

\bibitem{Wang2021}
Jiaqi Wang, Alexander Serb, Christos Papavassiliou, and Themistoklis Prodromakis.
\newblock Accounting for memristor {I-V} non-linearity in low power memristive amplifiers.
\newblock In {\em 2021 IEEE International Symposium on Circuits and Systems (ISCAS)}, pages 1--5, 2021.

\bibitem{Marcotte2023}
Frédéric Marcotte, Pierre-Antoine Mouny, Victor Yon, Gebremedhin~A. Dagnew, Bohdan Kulchytskyy, Sophie Rochette, Yann Beilliard, Dominique Drouin, and Pooya Ronagh.
\newblock A cryogenic memristive neural decoder for fault-tolerant quantum error correction, 2023.

\bibitem{Kuhlman2013}
Andreas~V. Kuhlmann, Julien Houel, Arne Ludwig, Lukas Greuter, Dirk Reuter, Andreas~D. Wieck, Martino Poggio, and Richard~J. Warburton.
\newblock Charge noise and spin noise in a semiconductor quantum device.
\newblock 9(9):570--575, July 2013.

\bibitem{Connors2019}
Elliot~J. Connors, JJ~Nelson, Haifeng Qiao, Lisa~F. Edge, and John~M. Nichol.
\newblock Low-frequency charge noise in {Si/SiGe} quantum dots.
\newblock {\em Phys. Rev. B}, 100:165305, Oct 2019.

\bibitem{mesoudy2021CMOS}
Abdelouadoud~El Mesoudy, Gw{\'{e}}naëlle Lamri, Raphaël Dawant, Javier Arias-Zapata, Pierre Gliech, Yann Beilliard, Serge Ecoffey, Andreas Ruediger, Fabien Alibart, and Dominique Drouin.
\newblock Fully {CMOS}-compatible passive {TiO$_\textrm{2}$}-based memristor crossbars for in-memory computing.
\newblock {\em Microelectronic Engineering}, 255:111706, February 2022.

\bibitem{Palmisano2001}
G.~Palmisano, G.~Palumbo, and S.~Pennisi.
\newblock Design procedure for two-stage cmos transconductance operational amplifiers: A tutorial.
\newblock {\em Analog Integrated Circuits and Signal Processing}, 27(3):177--187, 2001.

\bibitem{Boyn2014}
S.~Boyn, S.~Girod, V.~Garcia, S.~Fusil, S.~Xavier, C.~Deranlot, H.~Yamada, C.~Carrétéro, E.~Jacquet, M.~Bibes, A.~Barthélémy, and J.~Grollier.
\newblock High-performance ferroelectric memory based on fully patterned tunnel junctions.
\newblock {\em Applied Physics Letters}, 104(5), February 2014.

\bibitem{Hur2022}
Jae Hur, Chinsung Park, Gihun Choe, Prasanna~Venkatesan Ravindran, Asif~Islam Khan, and Shimeng Yu.
\newblock Characterizing hfo2-based ferroelectric tunnel junction in cryogenic temperature.
\newblock {\em IEEE Transactions on Electron Devices}, 69(10):5948–5951, October 2022.

\bibitem{cryo_ampli_le_guevel_2020}
L.~Le Guevel, G.~Billiot, B.~Cardoso Paz, M.~L.~V. Tagliaferri, S.~De Franceschi, R.~Maurand, M.~Cass{\'{e}}, M.~Zurita, M.~Sanquer, M.~Vinet, X.~Jehl, A.~G.~M. Jansen, and G.~Pillonnet.
\newblock Low-power transimpedance amplifier for cryogenic integration with quantum devices.
\newblock {\em Applied Physics Reviews}, 7(4):041407, December 2020.

\end{thebibliography}
